\documentclass[letterpaper,english,reprint, aps, pre,nofootinbib]{revtex4-1}
\usepackage[T1]{fontenc}
\usepackage[latin9]{inputenc}
\usepackage{color}
\usepackage{units}
\usepackage{amsmath}
\usepackage{amssymb}
\usepackage{graphicx}
\usepackage{esint}

\makeatletter

\pdfpageheight\paperheight
\pdfpagewidth\paperwidth

\makeatother

\usepackage{babel}
\begin{document}
\title{Controlled diffusion processes in an adiabatic model of a bouncing
ball }
\author{Valdir Barbosa da Silva Junior}
\author{Ana Laura Boscolo}
\author{Diogo Ricardo da Costa}
\author{Luiz Antonio Barreiro}
\email{luiz.a.barreiro@unesp.br}

\address{Institute of Geosciences and Exact Sciences, ~\linebreak{}
Physics Department, S�o Paulo State University (Unesp), CEP 13506-900,
Rio Claro, S�o Paulo, Brazil}
\date{\today}
\begin{abstract}
This study explores the integration of a diffusion control parameter
into the chaotic dynamics of a modified bouncing ball model. By extending
beyond simple elastic collisions, the model introduces elements that
affect the diffusive behavior of kinetic energy, offering insights
into the interplay between deterministic chaos and stochastic diffusion.
The research reinterprets the bouncing ball's interactions with the
surface as short-distance collisions that mimic random thermal fluctuations
of particles. This refined model reveals complex dynamics, highlighting
the synergistic effects between chaos and diffusion in shaping the
evolution of the system. 
\end{abstract}
\maketitle

\section{Introduction}

\textcolor{black}{Diffusion processes constitute a fundamental and
ubiquitous phenomenon in a wide range of scientific disciplines, playing
a central role in understanding the transport and spreading of matter,
energy, and information. Rooted in statistical physics and stochastic
processes, the concept of diffusion emerged from early observations
of random particle motion in fluids during the early 19th century.
Since then, the study of diffusion has evolved into a sophisticated
and interdisciplinary field with applications spanning physics, chemistry,
biology, engineering, economics, sociology, and beyond \citep{Cussler2009,Saxton2001,Geisel,Szymanski,Crank1981}. }

\textcolor{black}{From a mathematical standpoint, diffusion processes
are characterized by well-defined diffusion equations, such as the
classic Fick's laws of diffusion, the heat equation, and the stochastic
differential equations employed in stochastic processes \citep{Zwanzig2001}.
These equations offer a quantitative framework to model and analyze
the evolution of diffusing species in a myriad of contexts, ranging
from the diffusion of chemical substances in a liquid medium to the
diffusion of innovations in social networks. }

\textcolor{black}{The study of diffusion processes has yielded profound
insights and far-reaching applications. In physics, diffusion phenomena
are central to understanding the behavior of gases, liquids, and solids,
as well as critical to the understanding of mass and energy transport
in diverse systems \citep{Cussler2009}. In chemistry, diffusion plays
a pivotal role in chemical reactions, diffusion-limited processes,
and the dynamics of biological systems at the molecular level \citep{Atkins2018}.
Moreover, in biology, diffusion is instrumental in understanding cellular
transport processes, nerve impulse propagation, and the movement of
molecules across cellular membranes \citep{Alberts}. }

Building upon the scenarios previously described, this study integrates
a control parameter of diffusion processes into the chaotic dynamics
of a modified bouncing ball model. This refined extension goes beyond
simple collisions, introducing elements that influence the diffusive
behavior in kinetic energy. By examining the interplay between deterministic
chaos and diffusion, we aim to elucidate the synergistic effects that
shape the evolution of the bouncing ball system. This study investigates
diffusive processes within a modified bouncer model, wherein the elastic
collisions of the bouncing ball with the surface are reinterpreted
as short-distance interactions designed to emulate the random thermal
fluctuations of particles. The combination of deterministic chaos
and stochastic diffusion leads to rich dynamics where the interplay
between these two mechanisms can be observed.

This paper is structured as follows: Sec. II introduces the model
and outlines the mapping details. Sec. III explores the different
types of diffusive processes observed in the problem, as well as the
transitions between them. In Sec. IV, we analyze the correlation between
the phase space and the distinct diffusive behaviors of the particle.
Finally, Sec. V presents the conclusions and discusses future perspectives.

\section{The model}

In the current study, analogous to Boltzmann\textquoteright s approach
to understand Maxwell distribution \citep{Pathria2022}, we examine
particles moving under the influence of a gravitational field, but
we incorporate inelastic collisions with the ground. In this context,
\textquotedbl inelastic\textquotedbl{} denotes that a particle can
either lose or gain kinetic energy due to the thermal fluctuations
of the ground. 

At the molecular level, the interaction of a molecule with a surface
is understood as a short-range repulsive interaction between dipoles
present on both the surface and the molecule itself \citep{Huang1987}.
This interaction results in a variation in the particle's velocity,
which can be described by the following mapping: 
\begin{equation}
v_{n+1}=\gamma_{n+1}v_{n},\label{veloc1}
\end{equation}
where $n$ represents the $n$-th collision and the variable $v_{i}$
will consistently denote the speed immediately following the $i$-th
collision. The $\gamma_{n+1}$ factor, which controls the energy exchange
between the particle and the floor during the $(n+1)$-th collision,
is modeled as
\begin{equation}
\gamma_{n+1}=\gamma\left[1+\frac{\zeta\sin(\omega t_{n+1})}{v_{n}^{z}}\right].\label{gamman1}
\end{equation}
The term involving the sine function represents the thermal fluctuation
with frequency $\omega$ at time $t_{n+1}=t_{n}+\delta t_{n,n+1}$,
where $t_{n}$ is the instant of the $n$-th collision and $\delta t_{n,n+1}$
is defined as the travel time the particle spends between the $n$-th
collision and the $(n+1)$-th collision. The term $v_{n}^{z}$ will
be referred to as the \textit{adiabatic factor}, as the contribution
of thermal fluctuation diminishes with increasing velocity. The exponent
$z$ has an additional impact on this connection. As we shall see
in the numerical simulations, tiny modifications in the $z$ parameter
result in significant changes in system behavior. The factor $\zeta$
denotes the amplitude or intensity of the thermal fluctuation. If
$\zeta=0$, the parameter $\gamma_{n}$ reduces to a constant value
$\gamma$. Considering this case, the constant $\gamma$ determines
the nature of the collision: when $\gamma=1$, the collision is elastic,
preserving kinetic energy; conversely, when $\gamma=0$, the collision
results in the complete dissipation of kinetic energy. In general
case, any value between 0 and 1 is allowed. This article exclusively
examined the scenario where $\gamma$ is equal to 1.

To utilize Eq (\ref{veloc1}), it is necessary to calculate the time
$t_{n+1}$ based on the velocity $v_{n}$ and the time $t_{n}$. Given
that the particle moves under the influence of a gravitational field
$g$, the time interval between two successive collisions with the
ground is equal to twice the time taken during the ascent, which can
be readily determined by
\begin{equation}
t_{n+1}-t_{n}=\delta t_{n,n+1}=\frac{2v_{n}}{g}.\label{timeInterval}
\end{equation}

By defining the dimensionless variables 
\[
\phi_{n}=\omega t_{n+1}\textrm{ and }V_{n}=\frac{\omega}{2g}v_{n},
\]
we find
\begin{equation}
\begin{split}V_{n+1} & =\gamma\left[V_{n}+K_{n}(V_{n},\bar{\zeta},z)\sin(\phi_{n})\right]\\
\phi_{n+1} & =\phi_{n}+V_{n+1}
\end{split}
,\label{Smap-1-1}
\end{equation}
where we have defined the function
\begin{equation}
K_{n}(V_{n},\bar{\zeta},z)=\bar{\zeta}\frac{2^{z}}{V_{n}^{z-1}}\label{Smap-2-1}
\end{equation}
and also the parameter $\bar{\zeta}=\zeta\,(\omega/g)^{z}$. Eqs (\ref{Smap-1-1})
are very similar to the Chirikov-Taylor map \citep{chirikov1979}
or, as it is better known, standard map. A new reparameterization
of the variables ($\phi_{n}\rightarrow\phi_{n+1}$and $V_{n}\rightarrow2V_{n}$
) leads the system to be equivalent to a simplified version of the
Fermi-Ulam model used by Lieberman and Lichtenberg \citep{Lichtenberg_1992}
to ease computational simulations, often referred to in the literature
as the simplified bouncer model \citep{Livorati}. When $\gamma=1$
and $z=1$, the system reduces to the standard map model. In the iterative
process, if the particle acquires a negative velocity, it is considered
absorbed by the surface. Subsequently, the particle is re-injected
with the same positive velocity. Another observation that can already
be made is that when the value of $z<1$, the exponent $z-1$ becomes
negative, causing the factor $V_{n}^{z-1}$ to move to the numerator.
Consequently, if the speed $V_{n}$ is small, the tendency is for
the speed values to decrease with each new iteration. 

\section{Diffusion process}

In the context of the bouncer model, researchers investigate various
aspects such as the scaling properties of Fermi acceleration, the
effects of dissipative forces, and the statistical properties of chaotic
orbits \citep{Chastaing2015,Liang2010,Boscolo2023}. It is primarily
used to study Fermi acceleration \citep{Fermi1949}, where a particle
can gain energy through repeated collisions with a moving boundary.

In the present model, it is possible to use an analytical argument
to predict the unlimited diffusion of energy through the Fermi acceleration
mechanism. Let us consider an ensemble of particles with initially
low velocities, which means an ensemble characterized by a low temperature,
but sufficiently high to neglect quantum effects. By squaring the
first equation in mapping (\ref{Smap-1-1}), we obtain
\begin{eqnarray*}
V_{n+1}^{2} & = & \gamma^{2}V_{n}^{2}+2\gamma^{2}K_{n}\sin(\phi_{n}+2V_{n})\\
 &  & +\gamma^{2}K_{n}^{2}\sin^{2}(\phi_{n}+2V_{n}).
\end{eqnarray*}
Making an ensemble average over $\phi\in[0,2\pi]$, we obtain
\[
\left\langle V_{n+1}^{2}\right\rangle =\gamma^{2}\left\langle V_{n}^{2}\right\rangle +\frac{1}{2}\gamma^{2}\left\langle K_{n}^{2}\right\rangle .
\]
For a sufficiently large $n$, we can express the relationship as
follows:
\begin{eqnarray*}
\frac{\partial\left\langle V_{n}^{2}\right\rangle }{\partial n} & \simeq & \frac{\left\langle V_{n+1}^{2}\right\rangle -\left\langle V_{n}^{2}\right\rangle }{(n+1)-n}\\
 & = & \left(\gamma^{2}-1\right)\left\langle V_{n}^{2}\right\rangle +\frac{1}{2^{1-z}}\gamma^{2}\bar{\zeta}^{2}\left\langle V_{n}^{2-2z}\right\rangle .
\end{eqnarray*}

To advance further, let us assume a Gaussian form for the distribution,
with an anomalous diffusion \citep{Cecconi2022,Boscolo2023}. Under
this assumption, we can set (see Appendix \ref{sec:Appendix:AnomalousDistribution})
$\left\langle V_{n}^{2-2z}\right\rangle \thickapprox\alpha\left\langle V_{n}^{2}\right\rangle ^{1-z}$
leading to a differential equation in $\left\langle V_{n}^{2}\right\rangle $.
The solution to this equation can then be expressed as{\small{}
\[
\left\langle V_{n}^{2}\right\rangle =\left\{ V_{0}^{2z}e^{z(\gamma^{2}-1)n}+\frac{\alpha\gamma^{2}\bar{\zeta}^{2}}{2^{1-z}(\gamma^{2}-1)}\left[e^{z(\gamma^{2}-1)n}-1\right]\right\} ^{\nicefrac{1}{z}}.
\]
}{\small\par}

Assuming that the collisions are approximately elastic, we can take
the limit as $\gamma$ approaches 1 by expanding $e^{z(\gamma^{2}-1)n}\approx1+z(\gamma^{2}-1)n$.
Additionally, we introduce the $\beta$ function as
\begin{equation}
\beta(z,\bar{\zeta})=\left(2^{z-1}z\right)\bar{\zeta}^{2}\alpha=\frac{z}{\sqrt{\pi}}\bar{\zeta}^{2}\varGamma\left(\frac{3}{2}-z\right).\label{alphabar}
\end{equation}
So the root mean square ($rms$) velocity, defined by $V_{rms}^{(z)}=\sqrt{\left\langle V_{n}^{2}\right\rangle }$,
is expressed as
\begin{equation}
V_{rms}^{(z)}=\left\{ V_{0}^{2z}+\beta\,n\right\} ^{\nicefrac{1}{2z}}\sim\beta^{\nicefrac{1}{2z}}\sqrt[2z]{n}.\label{V2-2}
\end{equation}

The result (\ref{V2-2}) suggests that the diffusion behavior in the
system is dependent on the value of the parameter $z$. Specifically,
when $z=1$, the system exhibits normal diffusion. However, for $z<1$,
the system displays superdiffusive behavior, while for $z>1$, subdiffusive
behavior is observed. In this context, normal diffusion corresponds
to the fact that the mean square velocity (MSV) is linearly proportional
to n. Superdiffusion is characterized by an MSV that increases faster
than linearly with time, indicative of long-range correlations or
persistent motion within the system. Conversely, subdiffusion is marked
by an MSV that grows slower than linearly with time, often arising
from stability regions or hindrance effects. Thus, the adiabatic parameter
$z$ serves as a critical control mechanism that dictates the transition
between subdiffusive, diffusive, and superdiffusive states in the
energy dynamics of the system. 

Therefore, understanding the role of the adiabatic parameter $z$
is essential for predicting and controlling the diffusive properties
of the system. This parameter effectively modulates the rate and nature
of energy transfer within the system, influencing the overall dynamic
behavior and allowing for precise manipulation of diffusion processes.
\begin{figure}
\centering{}\includegraphics[width=1\columnwidth]{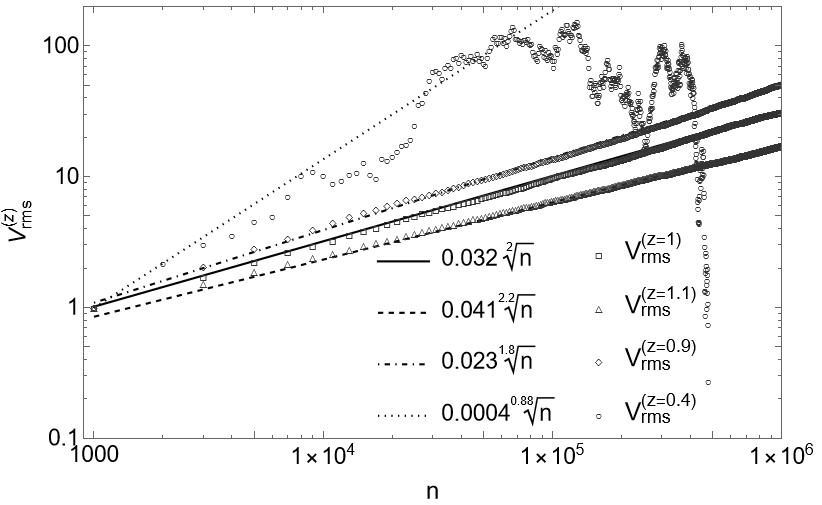}\caption{Vrms plots for 4 different values of $z$, considering an ensemble
of $10^{4}$ initial conditions varying $\phi$ in the interval $[0,2\pi]$
and evolving $10^{6}$ iterations. For a better comparison between
the graphs, a normalization was introduced such that $V_{rms}^{(z)}(1000)=1$
for any $z$ value. This normalization facilitates a more effective
comparison of the $V_{rms}^{(z)}$ behaviors across different diffusive
regimes. The iterations are conducted using $\gamma=1$,$\bar{\zeta}=5$,
$z=1.1,\,1.0,\,0.9$, and $0.4$. The solid, dashed, dot-dashed and
dotted lines represent the theoretical results represented by Eq (\ref{V2-2}).}
\label{FigVrms}
\end{figure}

To substantiate this behavior, we conducted numerical simulations
with $n=10^{6}$ iterations and $10^{4}$ different initial conditions,
for four different values of the parameter $z$: specifically, $z=1$,
$z=1.1$, $z=0.9$, and $z=0.4$. According to Eq. (\ref{V2-2}),
the anticipated results are as follows: for $z=1$, the result is$\sqrt[2z]{n}=n^{1/2}$;
for $z=1.1$, the result is $\sqrt[2z]{n}=n^{1/2.2}$; for $z=0.9$,
the result is $\sqrt[2z]{n}=n^{1/1.8}$; and for$z=0.4$, the result
is $\sqrt[2z]{n}=n^{1/0.8}$. The numerical results obtained using
Map (\ref{Smap-1-1}), with the values of the $\gamma$ and $\bar{\zeta}$
parameters set to 1 and 5, respectively, are illustrated in Figure
\ref{FigVrms}, alongside the power functions that fit the data. The
number of iterations ($n=10^{6}$ ) ensures that the simulations capture
the long-term behavior of the system, thereby providing a comprehensive
understanding of the diffusion processes. Additionally, using $10^{4}$
different initial conditions allows us to account for variability
and ensures that the results are not dependent on any specific starting
point, thereby enhancing the robustness and generalizability of our
findings. The numerical findings corroborate the theoretical results
for the initial three cases, $z=1,\,1.1$ and $0.9$. However, for
$z=0.4$, $V_{rms}^{(z)}$ initially exhibits diffusive behavior,
approximately aligning with the theoretically predicted result. Yet,
for larger iteration values $(n>10^{5})$, $V_{rms}^{(z)}$ abruptly
approaches zero, indicating a dissipative process.

\begin{figure}
\begin{centering}
\includegraphics[width=1\columnwidth]{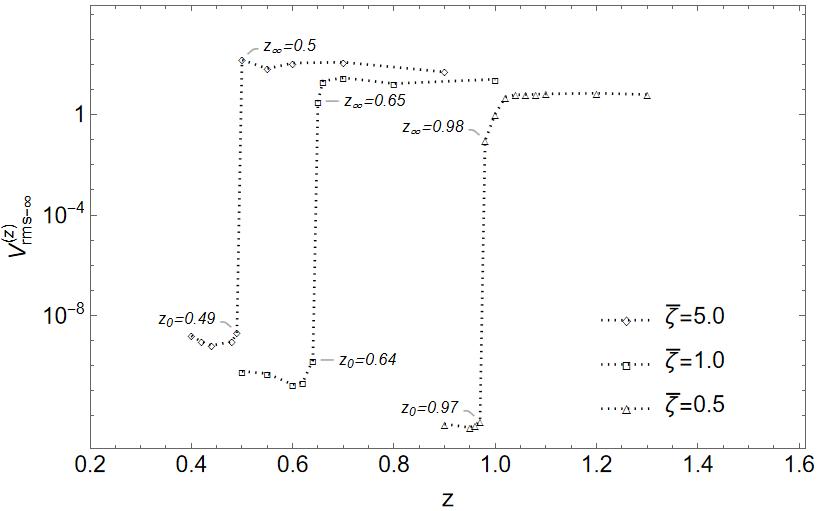}\caption{Graphs illustrating transition regions between the frozen phase and
the diffusive phase for three distinct values of zeta. The critical
values of z at which these transitions occur are emphasized. }
\label{Vthreshold}
\par\end{centering}
\end{figure}

To elucidate this final outcome, Figure \ref{Vthreshold} presents
a graph depicting $V_{rms}^{(z)}$ at the $10^{6}$th iteration, which
we shall hereafter refer to as $V_{rms-\infty}^{(z)}$, as a function
of $z$.

 It is clearly observed that, for a given $\bar{\zeta}$, there exists
a critical value for $z$, below which $V_{rms-\infty}^{(z)}$ tends
towards zero ($V_{rms-\infty}^{(z)}<10^{-8}$). This behavior can
be interpreted as a \textquotedbl freezing\textquotedbl{} of the
particle system, where the particles lose their dynamic motion and
the system reaches a static state. In the notation used in the graph,
$z_{\infty}$ represents the smallest value of $z$ that still permits
diffusive processes. Conversely, $z_{0}$ represents the threshold
value at which the system transitions into the frozen phase. The critical
value $z_{\infty}$ serves as a point distinguishing between diffusive
behavior and the frozen phase, thereby providing significant insight
into the underlying dynamics of the system. The graph illustrates
the range of $z$ values within which freezing occurs. For instance,
when $\bar{\zeta}=5$, the value of $z$ that still permits diffusive
states is $z_{\infty}=0.50$. If this value is decreased to $z_{0}=0.49$,
the system transitions to the frozen state. Thus, a mere change of
0.01 in the $z$ value significantly alters the system's behavior.
For $\bar{\zeta}=0.3$, the limiting value of $z_{\infty}$ for the
occurrence of diffusive states approaches 1. 

These results can be further explored by developing a phase diagram
that clarifies the relationship between the system's behavior and
the parameters $\bar{\zeta}$ and $z$. This process entails generating
a grid of discrete points, with $\bar{\zeta}$ values ranging from
0.2 to 5 and $z$ values extending from 0.2 to 1.2. For each unique
pair $(\bar{\zeta},z)$ within this grid, we compute the $V_{rms}^{(z)}$.
Inspired by Eq. (1), we model the growth of $V_{\text{rms}}$ as a
power-law function of $n$ (the number of iterations), expressed as
$V_{\text{rms}}=an^{b}$. The resultant $V_{rms}^{(z)}$ values are
subsequently categorized according to the exponent $b$, which provides
insight into the diffusion behavior of the system. Specifically, for
$b<0.49$, the system is classified as exhibiting subdiffusive behavior,
indicating slower-than-normal diffusion. In the range $0.49<b<0.51$,
the system demonstrates standard diffusive characteristics, where
the diffusion process is characterized by a linear relationship with
time. Conversely, for $b>0.51$, the system is classified as superdiffusive,
suggesting a diffusion process that occurs at a rate faster than that
of standard diffusion.

\begin{figure}

\begin{centering}
\includegraphics[width=0.9\columnwidth]{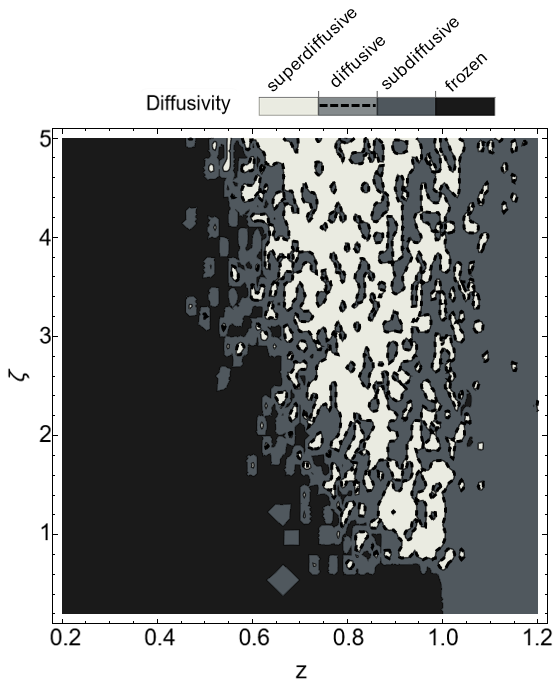}\caption{Phase diagram categorized by the power-law relationship $V_{\text{rms}}=an^{b}$.
For $b<0.49$, the process is identified as subdiffusive, while for
$b>0.51$, it is classified as superdiffusive. The boundary between
these two regimes, where $0.49<b<0.51$, corresponds to a normal diffusion
process. \label{FigDifusiv}}
\par\end{centering}
\end{figure}

These classifications and their corresponding behaviors are illustrated
in the accompanying Figure \ref{FigDifusiv}, which visually represents
the varying diffusion regimes as a function of the parameters $\zeta$
and $z$. 
Given that normal diffusive behavior is confined to a narrow range
of $b$ values, it is represented by dashed lines at the boundaries
between subdiffusive (light region) and superdiffusive regimes (gray
region). The dark region represents the frozen phase. 

For a complete analysis of this phase diagram, we will examine the
dimensionality of the normal diffusion boundary that separates subdiffusive
from superdiffusive behavior. For this we use the \textit{Box-Counting
Method}, also known as the \textit{Minkowski-Bouligand Dimension}
\citep{Peitgen2004}, is a common technique used to estimate the fractal
dimension of a set. This method involves covering the set with a grid
of boxes of size $\epsilon$ and determining how the number of boxes
$N(\epsilon)$ that contain part of the set scales as the box size
$\epsilon$ decreases.

The dimension $D$ is defined by the scaling relation:
\[
D=\lim_{\epsilon\to0}\frac{\log N(\epsilon)}{\log(1/\epsilon)}
\]
In practice, we compute $N(\epsilon)$ for several values of $\epsilon$
and estimate the slope of the log-log plot of $N(\epsilon)$ versus
$1/\epsilon$. This slope corresponds to the fractal dimension $D$.
For the set of points defining the normal diffusive boundary, we determine
the following dimensional value: $D=1.52439$. This result suggests
the fractal nature of the boundary in question.

Therefore, the adiabatic parameter $z$ plays a crucial role in determining
the dynamic behavior of the system. As $z$ varies, the system can
transition from superdiffusive processes to normal diffusion, and
ultimately to subdiffusive processes. We observe that within any diffusive
regime, there is an unlimited increase in energy, a phenomenon referred
to as Fermi acceleration. Furthermore, if the z parameter falls below
a certain threshold, the system undergoes a ``phase transition'',
resulting in a \textquotedbl frozen\textquotedbl{} state. This indicates
a rich structural complexity influenced significantly by variations
in the $z$ parameter. To gain a deeper understanding of these results,
it is essential to study the stability and chaos within this system.

\section{Chaotic properties}

The primary objective of this section is to examine the chaotic properties
of the mapping to understand the transitions between diffusion regimes
and the evolution of chaos during this process. Specifically, we aim
to analyze how changes in key parameters influence the system's behavior
and the structure of its phase space.

\begin{figure*}[!t]
\begin{centering}
\includegraphics[width=1\textwidth]{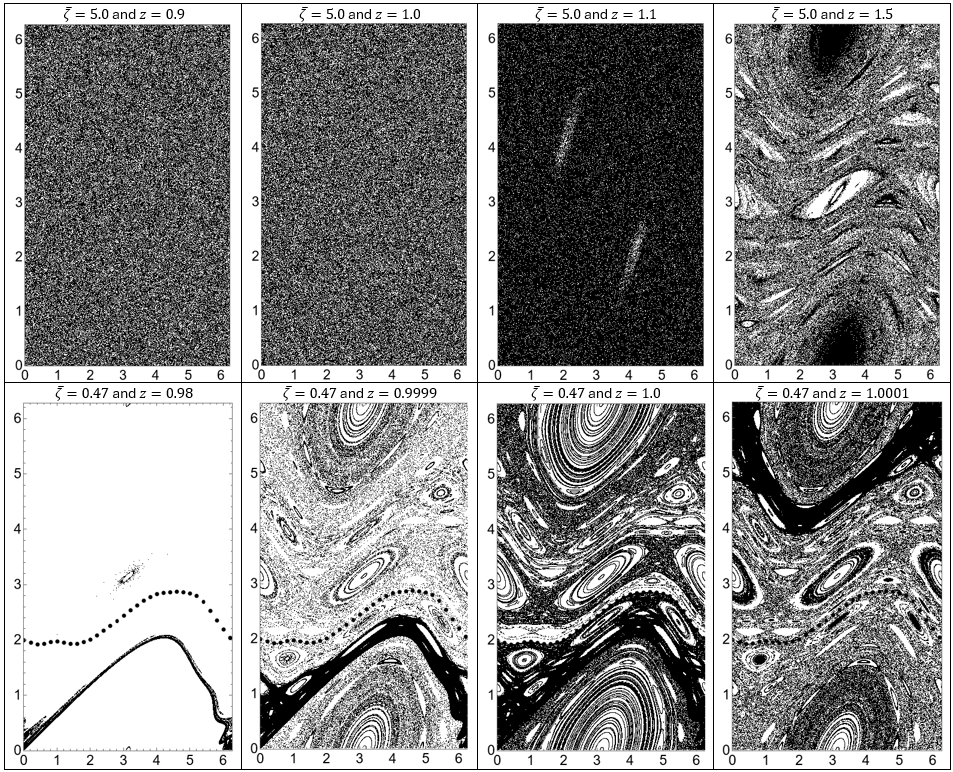}
\par\end{centering}
\caption{Phase spaces for various values of $\zeta$ and $z$. In all graphs,
the x-axis corresponds to $\phi$, while the y-axis corresponds to
$V$. The initial 200,000 data points are excluded to eliminate any
transient states. In the first row, phase spaces corresponding to
$\bar{\zeta}=5$ are shown, with $z$ varying from 0.9 to 1.5. The
second row illustrates additional phase spaces for $\bar{\zeta}=0.47$,
with $z$ exhibiting small variations in the range from 0.98 to 1.0001.
It is important to note that the phase spaces with $z=1$ represents
the established results for the standard map. }
\label{FigPhaseSpace}
\end{figure*}

\subsection{Phase space}

The phase space for map (\ref{Smap-1-1}) is dependent on the value
of the $\bar{\zeta}$ parameter as well as the adiabatic parameter
$z$. Variations in these parameters result in significant changes
to the phase space structure, which can provide insights into the
underlying dynamics of the system. By systematically altering the
$\zeta$ parameter and $z$, we can observe how the mapping transitions
between different diffusion regimes, offering a deeper understanding
of the interplay between order and chaos in the system. The resulting
phase space structures are presented in Figure \ref{FigPhaseSpace}.
The figure presents two graph lines, each characterized by a significantly
different $\bar{\zeta}$ value. The first row illustrates the phase
spaces for $\bar{\zeta}=5$ with $z=0.9$, $z=1.0$, $z=1.1$, and
$z=1.5$. The $\bar{\zeta}$ value and corresponding $z$ values are
selected to facilitate comparison with the $V_{rms}$ results depicted
in Figure \ref{FigVrms}. A significant difference between the phase
spaces for $z=0.9$ and $z=1.0$ is not readily observable. However,
it is important to note that the former represents a superdiffusive
process, while the latter corresponds to a diffusive process. The
distinction between these diffusive processes will be more precisely
identified by evaluating the evolution of the distance between two
points in phase space through the calculation of Lyapunov coefficients.
This method allows for a quantitative analysis of the system's sensitivity
to initial conditions, thereby providing deeper insight into the underlying
dynamics of the diffusive and superdiffusive processes. By examining
how these distances change over time, the Lyapunov coefficients serve
as a critical tool in differentiating the behavior and stability of
the processes in question. 

Diffusive processes predominantly occur for $\zeta=5$, while \textquotedbl freezing\textquotedbl{}
phenomena are observed in the second row for $z$ values below the
threshold $z=0.98$. In the phase spaces depicted in the first row,
the presence of chaos is evident, leading to superdiffusive processes
for $\ensuremath{z<1}$ (e.g., $z=0.9$) and diffusive processes for
$z=1$. For $z>1$, stable structures emerge within the phase space,
which are expected to retard the increase in particle velocity, resulting
in subdiffusive processes.

The second row presents the phase spaces for $\bar{\zeta}=0.47$ with
$z=0.98$, $z=0.9999$, $z=1.0$, and $z=1.0001$. The chosen value
for $\bar{\zeta}$ was inspired by the $K$ value for the standard
map that still exhibits the last spanning invariant curve. In the
standard map, the last spanning invariant curve is destroyed when
the value of $K$ exceeds the critical value $K_{c}=0.9716\ldots$.
In this scenario, local chaos, with $K<K_{c}$, transitions to global
chaos when $K\geqslant K_{c}$ and the last invariant spanning curve
is destroyed. In the model under consideration, the standard map is
recovered when $z=1$. Therefore, in accordance with Eq.(\ref{Smap-2-1}),
we obtain a critical value for $\bar{\zeta}$:
\[
\bar{\zeta}_{c}=\frac{K_{c}}{2}=0.4858...
\]
By selecting $\zeta=0.47$, it is ensured that when $z=1$, the invariant
spanning curve remains present in the phase space, as illustrated
in the third graph of the second row in Figure \ref{FigPhaseSpace}.
The sequence of black points represents the last spanning invariant
curve. It should be noted that the invariant curve is, in fact, a
continuous set of points, not a discrete one. This graphical representation
using a discrete set is employed to illustrate where the last invariant
curve would lie. In the case of the standard map, $z=1$, any initial
condition placed at any point on the line will evolve along the line.
However, with a variation of the $z$ value by one part in ten thousand,
the invariant curve is transformed into an invariant region. In this
context, the term \textquotedbl invariant region\textquotedbl{} refers
to the phenomenon where, if any point on the original invariant curve
is selected as the initial condition, this point evolves into a location
within the dark regions, characterized by higher point densities,
as highlighted in the graphs. Once the point reaches these regions,
it becomes trapped within them, a property that was tested for up
to $10^{8}$ iterations. So, in our simulations, we observed that
these trajectories did not exhibit the stickiness phenomenon. 

The system can be classified as a non-twist map, as it satisfies the
condition where $\phi_{n}$ and $V_{n}$ are related by the equation
$2^{z}\bar{\zeta}(z-1)\sin(\phi_{n})=V_{n}^{z}$, when $z\neq1$.
Consequently, we observe that
\[
\frac{\partial\phi_{n+1}}{\partial V_{n}}=\gamma\left(1+(1-z)\frac{2^{z}\bar{\zeta}}{V_{n}^{z}}\sin(\phi_{n})\right)=0
\]
violating the twist condition $\partial\phi_{n+1}/\partial V_{n}\neq0$
\citep{Reichl2004}, which is a crucial statement for KAM\textquoteright s
theorem applicability to the map. So, The absence of shear (or twist)
renders the system's behavior less predictable, frequently resulting
in complex dynamics, such as the formation of transport barriers,
as observed in our simulations. Hence, when $z$ deviates from 1,
the system undergoes a transition from a twist map to a non-twist
map.

It is important to note that when $z<1$ is selected, the invariant
curve is shifted to an invariant region below its original position,
while for $z>1$, the corresponding invariant region is located above
the initial position of the invariant curve. These observations can
be seen in the phase spaces presented in the bottom row of graphs
in Figure \ref{FigPhaseSpace}. The invariant curve is included in
all phase spaces to facilitate the observation of these phenomena.
However, it should be emphasized that in graphs where $z\neq1$, the
curve is no longer present. 

When the adiabatic factor $z$ is reduced, the invariant region diminishes
proportionally until reaching a threshold at $z=0.98$. For values
of $z$ below this threshold, a transition to the frozen state occurs,
resulting in the phase space becoming vacant.

\begin{figure*}[t]
\centering{}\includegraphics[width=1\textwidth]{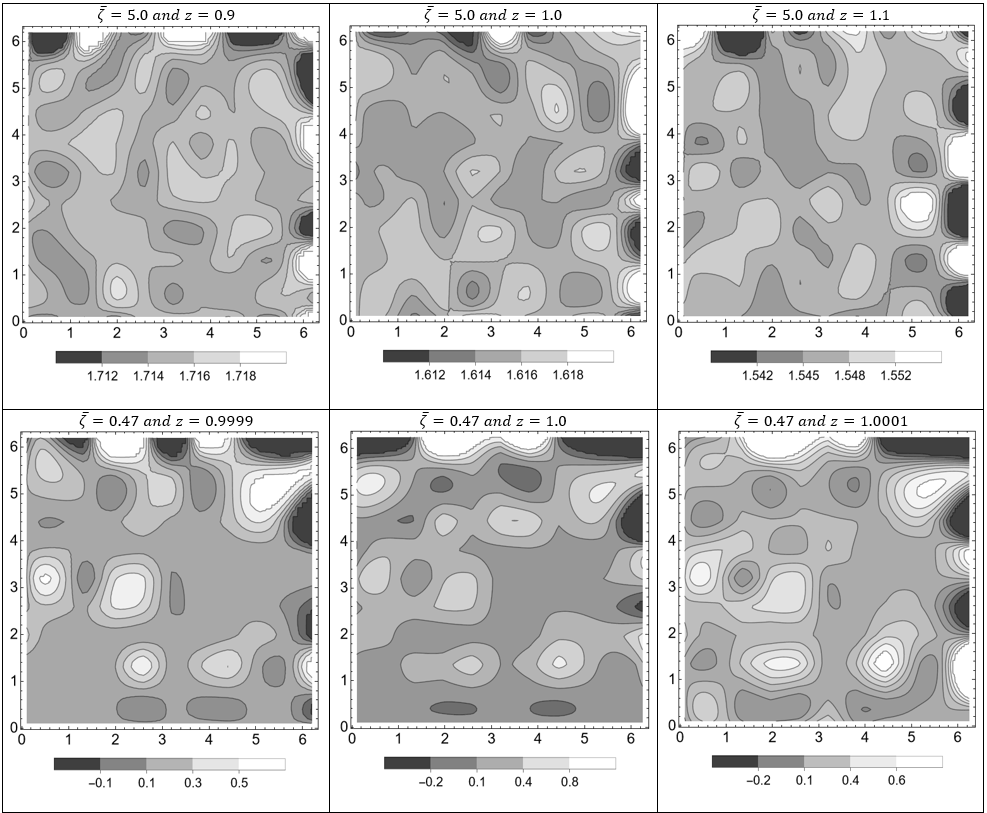}\caption{Variations in the values of the Lyapunov coefficients as a function
of the $\bar{\zeta}$ and $z$ parameters. In all graphs, the coordinate
axes depict the initial conditions. The initial phase $(\phi_{in})$
and initial velocity $(V_{in})$ are represented on the $x$-axis
and $y$-axis, respectively. The grayscale bar denotes the values
of the Lyapunov exponents.}
\label{LyapFig}
\end{figure*}

\subsection{Jacobian determinant and Lyapunov coefficients}

In the analysis of the dynamics of the mapping described in Eq (\ref{Smap-1-1}),
we observe that both chaotic and regular motion can coexist within
the phase space (Fig. \ref{FigPhaseSpace}). This coexistence creates
a scenario where there are significant variations and regions of local
instability along any given chaotic trajectory. Essentially, the system
does not follow a single predictable path but instead displays a complex
interplay between different types of motion. The Jacobian determinant
is given by
\[
\det[J]=1+(1-z)\frac{\bar{\zeta}}{V_{n}^{z}}\cos\left(2V_{n}+\phi_{n}\right).
\]
When $z\neq1,$ $\det[J]$ has the possibility of being different
from 1, suggesting that the system either expands or contracts the
local phase space volume, indicating the presence of chaotic behavior,
dissipation, or other complex dynamical phenomena.

The large variations in behavior are directly related to the system
switching between distinct states of motion. Specifically, we see
alternations between chaotic motion, where the system's behavior is
highly sensitive to initial conditions and appears disordered, and
quasiregular motion, where the system shows more predictable, though
not entirely regular patterns. This mix of behaviors within the same
system introduces a layer of complexity that makes the dynamics particularly
interesting and challenging to analyze.

To effectively characterize and understand these unique dynamic variations,
we analyze the Lyapunov coefficients along some trajectories, which
allows us to capture the intricate details of this modified standard
map dynamics and better understand the interplay between the various
diffusives regimes. The Lyapunov exponents for bidimensional systems
are defined as \citep{Eckmann1985}
\[
\lambda_{j}=\lim_{N\rightarrow\infty}\frac{1}{N}\ln\left|\Lambda_{j}^{N}\right|
\]
where $\Lambda_{j}^{N}$ are the eigenvalues of the matrix $M=\prod_{n=1}^{N}J(V_{n},\phi_{n})$
and $J$ is the Jacobian matrix evaluated over the orbit. The results
are illustrated in Figure \ref{LyapFig}. In constructing this figure,
the initial conditions were systematically chosen from an equally
spaced grid of points, where the variables $V_{\text{in}}$ and $\phi_{\text{in}}$
were sampled within the interval $[0,2\pi]$. Subsequently, an iterative
process was applied, extending over 1 million terms to ensure the
accuracy and stability of the results. 

In the first row of graphs, where $\zeta=5$, it is observed that
as the value of $z$ increases, the average value of the Lyapunov
exponents decreases. This trend indicates a transition in the system's
behavior: initially, it exhibits superdiffusive dynamics, then passes
through a regime of normal diffusion, and eventually becomes subdiffusive.
This observation aligns with the interpretation of Lyapunov exponents
as indicators of the divergence of trajectories in phase space. A
higher Lyapunov exponent corresponds to a faster rate of separation
between trajectories, which is associated with a more diffusive behavior
of the system.

\subsection{Transport properties}

\begin{figure}[h]

\begin{centering}
\includegraphics[width=0.9\columnwidth]{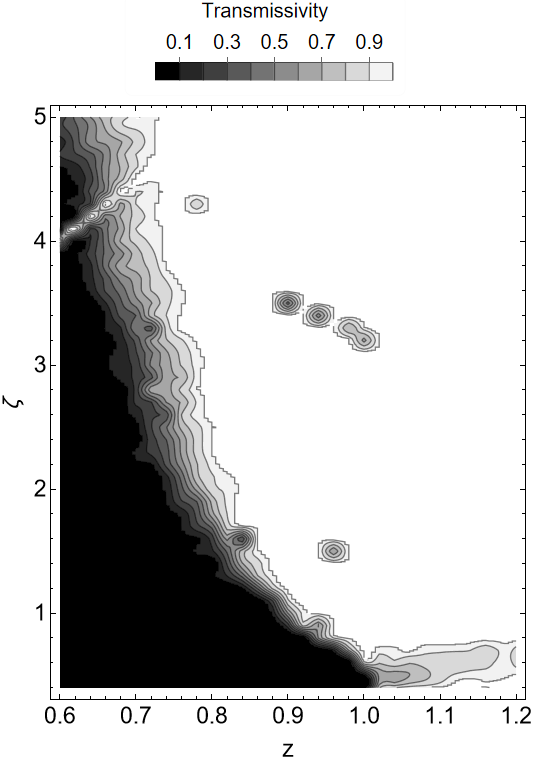}\caption{Transmissivity diagram. Fraction of orbits that reach or exceed the
position of the maximum value reached by the last invariant spanning
curve. The grayscale bar indicates the percentage of transmissivity.}
\label{FigTransm}
\par\end{centering}
\end{figure}

In the preceding sections, we examined the transition from a twist
map to a non-twist map. Nevertheless, even after the disruption of
the final invariant curve, an effective transport barrier may persist
in phase space, contingent upon the system's parameters. To account
for this phenomenon, we calculate the ratio of the number of orbits
that traverse the barrier to the total number of orbits, similarly
to what was done in Ref. \citep{Grime2023}. Initially, we examine
a set of conditions by fixing $V_{in}=1$ while varying $\phi_{in}$between
0.5 and 6.0. Through an iterative process consisting of $N=10^{5}$
steps, we determine the fraction of orbits that reach the points defined
by $V_{N}$=2.6, corresponding to the maximum value attained by the
last invariant spanning curve. The results are presented in Fig. \ref{FigTransm}.
The bar with a grayscale represents the fraction of orbits that pass
the barrier. In the black regions, this fraction of orbits is zero,
that is, the system has a total barrier. It is noteworthy that when
$z>1$, transmission occurs almost invariably, irrespective of the
value of $\zeta$. 
This behavior can also be understood in the phase space diagrams in
the second row of Figure \ref{FigPhaseSpace}, where it is apparent
that the region with a higher density of points progressively shifts
to higher values of $V$ as $z$ increases. This shift indicates that
orbits, which were previously obstructed, encounter decreasing resistance,
leading to a more extensive accessible region.

On the other hand, when $\bar{\zeta}$ assumes higher values, the
system experiences an increase in chaotic dynamics, which facilitates
greater transmissivity in phase space. In this context, transmissivity
refers to the ability of orbits to traverse through phase space regions
that might otherwise be obstructed. The increased chaos disrupts the
regularity of the system, allowing orbits to pass through regions
that would typically impede their progress.

The interface region, which lies between the zones of total blocking
and the onset of transmissivity, exhibits fluctuations. These fluctuations
are linked to variations in the local stability of the system, suggesting
that small changes in stability can lead to significant differences
in the behavior of orbits at this critical boundary. This behavior
underscores the complex interplay between stability, chaos, and orbit
dynamics within the system.

In graph of Figure \ref{FigTransm}, a clear qualitative similarity
to graph of Figure \ref{FigDifusiv} is observed, indicating that
the overall behavior of the system remains consistent across both
representations. Specifically, the freezing region in Figure \ref{FigDifusiv}
corresponds to the proportion of orbits that fail to overcome the
established barrier. This region visually demonstrates where the system
encounters resistance, leading to the trapping of orbits within a
specific range of values of $\bar{\zeta}$ and $z$.

In addition to the freezing region, the other regions in graph of
Figure \ref{FigTransm} exhibit analogous behavior to those in graph
of Figure \ref{FigDifusiv}. These regions reflect the varying capacities
of the orbits to either escape or remain confined by the system's
underlying constraints.

\section{Conclusions and outlook}

In this work, we analyzed the behavior of a boucing ball under the
influence of an adiabatic parameter, which governs the energy transfer
during each collision. The results demonstrate that the adiabatic
parameter not only controls the energy exchange, where the
particle may either gain or lose energy through an oscillatory interaction,
but also fundamentally alters the structure and dynamics of the particle's
phase space.

The variation of the adiabatic parameter introduces a diverse range
of dynamical regimes. For certain values, the particle can reach a
\textquotedbl frozen\textquotedbl{} state, characterized by zero
velocity ($V_{N}=0$), where the system appears to settle into a static
configuration. In contrast, other regions of the parameter space lead
to diffusive behaviors, with the particle exhibiting either subdiffusive
or superdiffusive motion. The boundary between these diffusive regimes
possesses fractal properties, reflecting the intricate, self-similar
structure of phase space and indicating a sensitive dependence on
initial conditions and parameter choices.

Moreover, the phase space is revealed to be mixed in nature, combining
regions of regular motion with chaotic zones. This is confirmed by
calculating the Lyapunov exponents, which indicate the presence of
both chaotic and regular dynamics. In regions where the Lyapunov exponents
are positive, the system exhibits chaotic behavior, characterized
by sensitivity to initial conditions and exponential divergence of
nearby trajectories. Conversely, regions with negative Lyapunov exponents
suggest stable, predictable motion, marking a clear distinction between
ordered and chaotic dynamics within the system. The presence of both
positive and negative Lyapunov exponents underscores the mixed-phase
nature of the system, with pockets of chaotic motion interspersed
with stable, regular regions.

Furthermore, the variation of the adiabatic parameter induces a transition
from a twist to a non-twist system. In classical twist systems, there
is a systematic, predictable progression of orbits in phase space,
but as the system shifts to a non-twist configuration, this order
breaks down. This transition suggests the breakdown of the KAM tori
and a fundamental change in the system\textquoteright s topological
properties, further contributing to the richness of the dynamical
behavior.

Finally, we evaluated the transmissivity of the particle across the
phase space. Our findings indicate that depending on the specific
choice of parameters, the phase space can either be fully accessible,
allowing the particle to explore the entire space, or partially accessible,
where certain regions become dynamically inaccessible. This restricted
access in certain parameter regimes introduces barriers in phase space,
confining the motion of the particle and limiting the range of possible
dynamical outcomes.

In conclusion, the interplay between the adiabatic parameter and the
particle\textquoteright s dynamics creates a complex and rich system,
where energy exchange mechanisms, diffusive behaviors, chaotic and
regular dynamics, and the twist-to-non-twist transition all contribute
to a highly structured phase space. The study provides insights into
how sensitive control parameters can significantly alter the global
and local dynamics of a system, with implications for understanding
similar phenomena in other physical systems governed by adiabatic
processes.

The ongoing research aims to further develop the study by establishing
a connection with thermodynamics. This is being achieved through the
formal definition and analysis of the system's entropy, which is expected
to offer a physical perspective on the model. By integrating the concept
of entropy, the study aims to enhance the theoretical framework and
offer new insights into the behavior of the system under thermodynamic
conditions. 

\section*{Acknowledgments}

The authors would like to thank Coordena��o de Aperfei�oamento de
Pessoal de N�vel Superior (Capes) for financial support.

\appendix

\section{Anomalous Distribution\label{sec:Appendix:AnomalousDistribution}}

Considering the Gaussian form of normal diffusion, with an anomalous
diffusion, we make a scaling hypothesis \citep{Cecconi2022} so that
we can express the anomalous distribution as
\begin{equation}
\Psi_{\mu}(V,n)=\sqrt{\frac{a}{\pi}}\frac{1}{n^{\mu}}\exp\left[-a\left(\frac{V}{n^{\mu}}\right)^{2}\right].\label{GaussForm-1}
\end{equation}
The associated moments are obtained as 
\begin{eqnarray*}
\left\langle V_{n}^{2}\right\rangle  & = & \intop_{-\infty}^{\infty}V^{2}\Psi_{\mu}(V,n)\,dV=\frac{n^{2\mu}}{2\sqrt{a^{2}}}.\\
\left\langle V_{n}^{2-2z}\right\rangle  & = & \intop_{-\infty}^{\infty}V^{2-2z}\Psi_{\mu}(V,n)\,dV\\
 & = & \frac{1}{\sqrt{a^{2-2z}\pi}}\varGamma\left(\frac{3-2z}{2}\right)\left(n^{2\mu}\right)^{1-z}.
\end{eqnarray*}
So we can conclude that
\[
\left\langle V_{n}^{2-2z}\right\rangle =\frac{2^{1-z}}{\sqrt{\pi}}\varGamma\left(\frac{3}{2}-z\right)\left\langle V_{n}^{2}\right\rangle ^{1-z}.
\]
Under this assumption, we can set $\left\langle V_{n}^{2-2z}\right\rangle =\alpha\left\langle V_{n}^{2}\right\rangle ^{1-z}$,
where $\alpha=\frac{2^{1-z}}{\sqrt{\pi}}\varGamma\left(\frac{3}{2}-z\right)$.

\section*{\textemdash \textemdash \textemdash \textemdash \textemdash \textendash{}}

\bibliographystyle{apsrev}
\addcontentsline{toc}{section}{\refname}\bibliography{references}

\end{document}